\documentclass[10pt,aps,twocolumn,prd,superscriptaddress,noshowpacs,nofootinbib,noshowkeys,floatfix]{revtex4}
\usepackage[dvips]{graphics,graphicx}
\usepackage{amsmath, amssymb}
\usepackage{multirow}
\usepackage{longtable}
\usepackage{color}
\usepackage[normalem]{ulem}  

\usepackage{hyperref}

\newcommand{\PbPb}         {\mbox{Pb--Pb}}
\newcommand{\pPb}          {\mbox{p--Pb}}

\begin{document}

\title{ Multiplicity dependence of light flavour hadron production at LHC energies in the strangeness canonical suppression picture}
\date{\today}
\author{V. Vislavicius}
\thanks{Vytautas.Vislavicius@cern.ch}
\affiliation{Division of Experimental High Energy Physics, University of Lund, Lund, Sweden}
\author{A. Kalweit}
\thanks{Alexander.Philipp.Kalweit@cern.ch}
\affiliation{European Organization for Nuclear Research (CERN), Geneva, Switzerland}

\begin{abstract}
We present an analysis of data on light flavour hadron production as function of event multiplicity at LHC energies measured by the ALICE collaboration. The strangeness-canonical approach within the framework of the THERMUS statistical hadronisation model is used for a simultaneous description of pp, p-Pb, and Pb-Pb collisions. The rapidity window dependence of the strangeness correlation volume is addressed and a value of $\Delta y = 1.33 \pm 0.28$ is found. With the exception of the $\phi$-meson, an excellent description of the experimental data is found.
\end{abstract}
\maketitle

\section{Introduction} 
The measured abundances of hadrons produced in heavy-ion collisions have been successfully described by statistical hadronization models over a wide range of energies~\cite{Andronic:2011yq,Cleymans:1998fq,Andronic:2005yp}. Thermal-statistical model calculations for central ultra-relativistic heavy-ion collisions are typically carried out in the grand-canonical ensemble. However, a grand-canonical description of particle production is only applicable if the volume $V=R^{3}$ of the system is large enough and as a rule of thumb one needs $VT^{3} > 1$ for a grand-canonical description to hold \cite{Hagedorn:1984uy,Rafelski:1980gk}. It must be noted, that this condition must be fulfilled for each conserved charge separately.
Several attempts were made to extend the picture of statistical hadronization to smaller systems such as pp or even e$^{+}$e$^{-}$ collisions~\cite{Redlich:2009xx,Becattini:1996gy,Kraus:2008fh}. While particle ratios of non-strange particles are observed to be very similar in small and large systems, the production of strangeness appears to be significantly suppressed in smaller systems. The data presented in~\cite{Adam:2016emw} shows for the first time, that there is a continuous increase of strangeness production with increasing event multiplicity across various collision systems. In the strangeness canonical approach, it is assumed that the total amount of strange hadrons in the volume is small with respect to non-strange hadrons. Thus the conservation of strangeness is guaranteed explicitly and not only on average while the bulk of the particles is still described in the grand-canonical ensemble. For further details on this approach, we refer for instance to~\cite{BraunMunzinger:2003zd,BraunMunzinger:2001as,Hamieh:2000tk,Cleymans:2004bf,Kraus:2007hf}. The study presented here utilises the implementation in the THERMUS code~\cite{Wheaton:2004qb}. The experimental data is taken from~\cite{Adam:2016emw,Abelev:2013haa,Adam:2016bpr,Adam:2015vsf,Abelev:2014uua,Abelev:2013xaa,Abelev:2013vea,ABELEV:2013zaa}.

\section{Correlation volume for strangeness production} 
Previous studies based on THERMUS were targeted to describe the evolution of multi-strange particle production in p-Pb collisions as a function of event multiplicity~\cite{Adam:2015vsf}. In this case, the volume for particle production was chosen such that the charged pion multiplicity d$N_{\pi}$/d$y$ at midrapidity corresponding to a window of $y = \pm 0.5$ units was correctly described by the model. This approach is equivalent to calculating strangeness suppression for a  system whose total extension only corresponds to one unit of rapidity. Consequently, the model describes qualitatively the suppression pattern, but overestimates the reduction of strangeness at small multiplicities. The discrepancy increases even further if this approach is extended to the even smaller multiplicities which are covered in pp collisions.

\begin{figure*}[t!]
  \begin{flushleft}
    \center{\includegraphics[width=\textwidth]{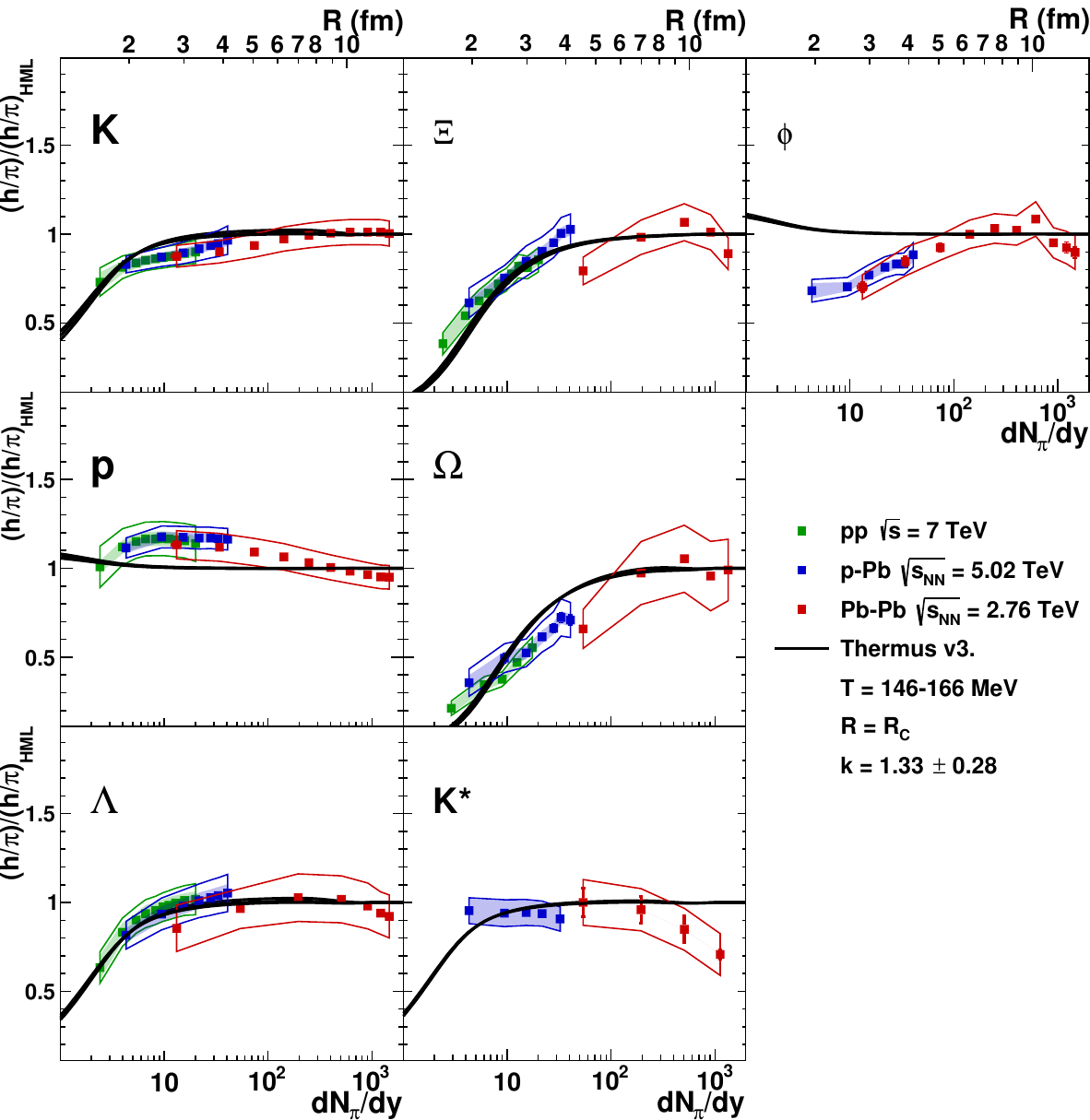}}
    \end{flushleft}
    \caption{(color online) Ratios of several particle species measured by the ALICE collaboration as a function of the midrapidity pion yields for pp, \pPb\ and \PbPb\ colliding systems compared to the THERMUS strangeness canonical suppression model prediction (black line), in which only the system size is varied.  All values except for the ${\rm K^{0*}}$ are normalised to the high multiplicity limit (see text for details). Note that $2\cdot{\rm K^{0}_{s}}$ are used for kaons in pp collisions, while ${\rm K^{\pm}}$ are used for p-Pb and \mbox{Pb-Pb} collisions. The upper axis shows the radius $R$ of the correlation volume $V=R^{3}$ which corresponds to the predicted particle ratios. The width of the model prediction line corresponds to a variation of the chemical freeze-out temperature between 146 MeV and 166 MeV. The data is taken from~\cite{Adam:2016emw,Abelev:2013haa,Adam:2016bpr,Adam:2015vsf,Abelev:2014uua,Abelev:2013xaa,Abelev:2013vea,ABELEV:2013zaa}.}
    \label{fig:SHM_Suppression}
\end{figure*}

In the study presented here, a similar approach as in~\cite{Adam:2015vsf} is followed: the strangeness saturation parameter is fixed to $\gamma_S = 1$, the chemical potentials $\mu$ of baryon and electric charge quantum number are set to zero, and the chemical freeze-out temperature $T_{ch}$ is varied from 146 to 166~MeV. Ratios of the production yields to pions are investigated for several particle species. In order to cancel the influence of the freeze-out temperature and to isolate the volume dependence, all ratios except for K$^{0*}$ are normalised to the high multiplicity limit, i.e. the grand-canonical saturation value for the model and the mean ratio in the 0-60\% most central Pb-Pb collisions for the data.
As the production rates of short-lived resonances in central heavy-ion collisions might be reduced by re-scattering effects in the hadronic phase~\cite{Abelev:2014uua,Adam:2016bpr}, the values for K$^{0*}$ were normalised to the most peripheral bin in Pb-Pb collisions. The resulting double ratios are shown in Fig.~\ref{fig:SHM_Suppression}.

In contrast to~\cite{Adam:2015vsf}, the total volume $V$ of the fireball was determined differently. While the strangeness conservation is also imposed to be of the size of the fireball ($V=V_C$), its absolute magnitude is not fixed to reproduce the pion multiplicity in a window of $y = \pm 0.5$ at midrapidity. Instead, it is fixed to reproduce a pion multiplicity which is a factor $k$ larger and thus corresponds to a larger rapidity window assuming a flat dependence of particle production as a function of rapidity. The same factor of $k$ was used for all particles and multiplicities. In practice, $k$ corresponds to a constant scaling factor of d$N_{\pi}$/d$y$ (the x-axis in Fig.~\ref{fig:SHM_Suppression}) and this feature is used for its numerical determination: the exact value of $k$ was optimised in a one parameter fit of the functions describing the evolution of the double ratios as a function of d$N_{\pi}$/d$y$ to the experimental data.  For the determination of the systematic uncertainty on the value of $k$, an alternative normalisation scheme for data was applied (normalisation to the highest available centrality bin) and the procedure for the determination of $k$ was repeated. A value of $k=1.33 \pm 0.28$ is obtained corresponding to a rapidity window of $y = \pm 0.66$. The results thus indicate that the total correlation volume for strangeness production extends over about 1.33 units in rapidity. In other words, strangeness as a conserved quantity in QCD can be effectively equilibrated over this distance in the system. Similar values can be obtained from purely theoretical considerations on causality constraints~\cite{Castorina:2013mba}. Furthermore, the size of the correlation window is also compatible with similar estimates for charm production~\cite{Andronic:2003zv,Andronic:2006ky}. We also note that this value is smaller than the plateau in the rapidity distribution at LHC energies which extends typically over three to four units \cite{Adam:2015gka,Abbas:2013bpa} and is thus meaningful from a physical point of view.

\section{Comparison to experimental data and conclusions.} As shown in Fig.~\ref{fig:SHM_Suppression} this approach allows for an excellent qualitative description of the particle ratios as a function of event multiplicity. They can be naturally described within in the framework of strangeness canonical suppression. The deviations observed for the K$^{0*}$ meson in central Pb-Pb collisions can be ascribed to the aforementioned re-scattering effects in the hadronic phase~\cite{Knospe:2015nva}. Furthermore, differences between the model and the data in the most peripheral Pb-Pb collisions in case of $\Omega$ and $\Xi$ can be potentially reduced with core-corona corrections~\cite{Aichelin:2008mi}.

From a quantitative point of view, essentially all data points can be described within 1-2 standard deviations. A potentially different trend is only observed for the $\phi$-meson for which -- as a strangeness-neutral particle -- a flat dependence as a function of event multiplicity is expected from the model, but which shows a rising and falling trend in data. Future experimental data will be needed in order to clarify the significance of this deviation. It must be noted though, that the $\phi$-meson also deviates from a common blast-wave fit to other light flavour hadrons in peripheral Pb-Pb collisions indicating an out-of-equilibrium production except for most central Pb-Pb collisions~\cite{Abelev:2014uua}.

Independent of the experimental precision and possible higher order effects in the particle production mechanisms, we find that the strangeness canonical approach can reproduce the multiplicity dependence of all measured light flavour hadrons across various collision systems on the 10-20\% level.

\acknowledgments
The authors would like to thank A. Andronic, F. Antinori, F. Bellini, P. Braun-Munzinger, D. Chinellato, M. Floris, and J. Schukraft for stimulating discussions and especially H. Oeschler for pointing out the rapidity window dependence of the strangeness correlation volume.

\bibliographystyle{utphys} 	
\bibliography{StrangenessCanonical}

\end{document}